\RequirePackage{fix-cm}
\documentclass[smallextended,natbib]{svjour3}       
\smartqed  
\usepackage{graphicx}
\usepackage{mathptmx}      
\journalname{Mind \& Society}
\begin{document}

\title{On the dynamics emerging from pandemics and infodemics 
}

\titlerunning{Pandemics and infodemics}        

\author{Stephan Leitner     
}
\authorrunning{Stephan Leitner} 
\institute{Stephan Leitner \at
	University of Klagenfurt\\
              Universit\"atsstra{\ss}e 65-67 \\
              9020 Klagenfurt, Austria\\
              \email{stephan.leitner@aau.at}           
}


\maketitle

\begin{abstract}
This position paper discusses emerging behavioral, social, and economic dynamics related to the COVID-19 pandemic and puts particular emphasis on two emerging issues: First, delayed effects (or second strikes) of pandemics caused by dread risk effects are discussed whereby two factors which might influence the existence of such effects are identified, namely the accessibility of (mis-)information and the effects of policy decisions on adaptive behavior. Second, the issue of individual preparedness to hazardous events is discussed. As events such as the COVID-19 pandemic unfolds complex behavioral patterns which are hard to predict, sophisticated models which account for behavioral, social, and economic dynamics are required to assess the effectivity and efficiency of decision-making. 

\keywords{COVID-19 \and delayed effects of pandemics \and individual preparedness \and policy making \and adaptive society \and social dynamics }
\end{abstract}

\section{Emergent dynamics related to the COVID-19 pandemic}

The COVID-19 pandemic poses extreme and severe challenges to the society: There is a dramatic loss of lives worldwide, we experience and also expect for the future a multiplicity of challenges for economic systems, individuals are confronted with new and often very demanding situations, and public and private institutions are challenged to decide upon draconian measures under high time-pressure  \citep{Hopkins2020,Atkeson2020,Xiang2020}. 

This paper discusses emergent issues which appear to become particularly relevant in the context of the COVID-19 pandemic: Section \ref{sec:2} focuses on delayed effects of pandemics and Section \ref{sec:3} discusses adaptive societies and preparedness in the context of hazardous events.  
The understanding of the dynamics which unfold during and in the direct aftermath of a pandemic are key to effective and efficient management decisions. In additon, the insights into the dynamics of pandemics and infodemics\footnote{A infodemic is referred to as the rapid diffusion of misinformation that accompanies a pandemic \citep{Zarocostas2020,Vaezi2020}.} can be employed for the good of the society in a long-term perspective as they might help minimize unwanted (and often delayed) effects.

\section{Delayed effects and their consequences for economic and social systems}
\label{sec:2}

The temporal profile of hazardous events significantly affects how the related risks are perceived: Individuals are more averse to events in which a large number of people are harmed or killed in a short period in time, compared to events which have similar consequences but span over a longer time-period \citep{Slovic1987,Bodemer2013,Ayton2019}. The COVID-19 pandemic, a low-probability and high-consequence event, is of the former type; individuals are very likely to perceive it as a  \textit{dread risk} \citep{Gerhold2020}. This perception is likely to affect individual behavior, which might unfold rather complex dynamics: \cite{Gigerenzer2004}, for example, shows that such dread risks affect individual decision-making behavior not just in the short but also in the long run. For the aftermath of the 9/11 attacks he finds that perception of the terrorist attack has caused a change in individual transport behavior and, in consequence, has led to an increase in traffic fatalities beyond what would have been expected without the change in individual behavior. It is likely that similar behavioral patterns will emerge in the aftermath of the COVID-19 pandemic.

These effects are likely to be shaped by two factors: The individual mind and structure of the environment \citep{Gaissmaier2012,Simon1990}. 
The former is affected by the perception of the COVID-19 pandemic as a dread risk with changes in behavioral patterns lurking \citep{Gigerenzer2004}. The latter also affects individual behavior as it determines the boundaries for adaptivity \citep{Simon1990}. The structure of the environment is, amongst others, shaped by policy decisions. These two influencing factors are discussed in the following subsections.

\subsection{The role of accessibility of (mis-)information}

Risks are made up of a multiplicity of qualitative and subjective attributes \citep{Jenkin2006,Slovic1981,Xu2020}.\footnote{ This is particularly true for non-experts, one common explanation is that experts put more emphasis on probabilities of harm or injury \citep{Sjoberg1999}. For further discussions related to factors driving differences in risk-perception between experts and non-experts and factors affecting individual risk-assessment see  \cite{Sjoberg1999a} and \cite{Bodemer2013}, respectively.} However, risk-perception is also a socio-cultural phenomenon: It is affected by the structure of networks between individuals (e.g., social networks, organizations) and the resulting world views \citep{Gore2009,Sjoberg2000,Marris1998}. How individuals perceive risks is crucially affected by the information accessible to them \citep{Huurne2008}, whereby accessibility of information and resulting opinions about factors influencing risk-perception are often shaped by one's networks \citep{Burt1987,Scherer2003,Grimm2020}.\footnote{For an extensive review of factors influencing information access see \cite{McCreadie1999} and \cite{McCreadie1999a}.}

Phenomena such as the \textit{infodemic}, which accompanies the current pandemic, adds complexity to the the current situation: There is a chance that mis-information leads to unwanted effects in individual behavior so that the speed at which the virus spreads increases: \cite{Cinelli2020} argue that rumors about lockdowns in northern Italy led to overcrowded trains and airports, which, in consequence, increased the speed at which the virus spreads. Besides direct effects on behavior there might be indirect and delayed effects of an infodemic, as the spread of mis-information might increase the perception of the COVID-19 pandemic as a dread risk \citep[][]{Zarocostas2020}. It is, thus, plausible to assume that there are non-trivial interactions among behavioral implications induced by the COVID-19 pandemic and the accompanying infodemic which might result in complex dynamics: If not understood properly, mutually reinforcing dynamics with unprecedented consequences might unfold. These dynamics might be even reinforced by the algorithms employed by information platforms, which are usually designed to respond to the individual information-seeking behavior \citep{Budak2016}. It is, therefore, of ultimate importance to understand how people select information sources during hazardous events and how information dynamics interfere with risk perception and behavioral dynamics \citep{Cinelli2020,Sharot2020}.

\subsection{The role of policy-making}

The structure of the environment is the second driving force behind delayed effects of the COVID-19 pandemic which is discussed in this paper as, following \cite{Simon1990}, it defines the limits of the adaptation. \cite{Gaissmaier2012}, for example, regard the availability of driving opportunities as a main factor that contributed to the change in behavioral patterns\footnote{In terms of a shift from flying to driving \citep{Gigerenzer2004}.} in the aftermath of the 9/11 attacks which led to more fatalities than would have been expected without the change in behavior \citep[see also][]{Gigerenzer2004}. Similarly, \cite{LopezRousseau2005} analyzed traffic patterns in the aftermath of terrorist train attacks in Spain in 2011: They observed that train travel had decreased in the months following the attacks, the amount of other traffic, however, had not increased.\footnote{\cite{LopezRousseau2005} argues that Spaniards might increased carpooling, decided to travel by bus or might have reduced travel alltogether.} Aside from cultural reasons, \cite{LopezRousseau2005} traces the differences back to political decisions and social factors: While the boundaries for individual adaptation after the 9/11 attacks were set by shutting down all flight traffic, social factors caused a significant increase in train traffic on the days after the attack in Spain as demonstrations related to the attacks were organized across the country. 

Thus, in the aftermath of hazardous events the environment appears to be a main factor to affect (and to set the boundaries for) the adaptation of individual behavior, whereby it appears to be shaped, amongst others, by policy decisions.\footnote{There are, of course, other factors which affect the adaptation of individual behavior such as personal attitudes and norms \citep{Montano2015}.} This is why policy-makers would be well advised to consider the dynamics resulting from the interaction of decisions related to the aftermath of the COVID-19 pandemic with processes of individual adaptation in their policy decisions. Currently, there is a multiplicity of (often draconic) measures taken by different governments which are decided upon under time pressure and under limited information: It is, therefore, likely that policy-makers are not fully aware of the impact of their decisions \citep{Elsenbroich2020}.
This might be explained by the fact that the models currently employed hardly consider the full range of social and behavioral complexity \citep{Squazzoni2020}: They are well-suited for short-term policy-making which aims at reducing the speed at which the virus spreads. In order to provide proper policy advice for long-term decisions, however, extended models need to be developed in order to avoid poorly conceived policies which strike back through delayed behavioral effects. 

\section{Preparedness and adaptive societies}
\label{sec:3}

The second emerging topic discussed in this paper covers issues related to the preparedness of societies.\footnote{Preparedness is related to the concept of resilient societies, i.e., societies which are able to cope with external stresses as a result of social, political, and/or environmental change \citep{Adger2000}.} This is an issue of ultimate interest: \cite{Oppenheim2019}, for example, highlight that despite significant investment, many countries are not able to manage virus outbreaks. Preparedness, however, is not just a necessary feature at the macroscopic level but also needs to be analyzed at the level of the individual \citep{Kaser2019}. There exist strong interrelations between the two levels: \cite{Lim2013}, for example, found that institutional preparedness to disasters is positively affected by the individual preparedness of supervisors, while individual  preparedness, amongst others, appears to be driven by the a person's experience with similar situations, the preparedness of colleagues and the family, and training. In order to increase the preparedness of a society, a multiplicity of issues has to be considered, such as communication and coordination infrastructure, health-care infrastructure, logistics to name but a few \citep{Madad2020,Oppenheim2019,Gupta2018}. However, soft factors which influence behavior at the individual level, such as strategies for information retrieval, also need to be taken into account \citep{Misuraca2018}.

Issues related to preparedness can be analyzed through the lens of complexity science: Societies are regarded as adaptive systems which consist of multiple human decision-makers and self-organization refers to a process in which the society, for example, creates an order or a structure and  assigns roles, tasks or capabilities to be acquired by their members  \citep{Odum1988,DiMarzoSerugendo2005}. What appears to be particularly interesting in this context is how societies can be guided in their process of self-organization so that their resilience increases. 
There is enormous progress in the fields of engineering and information technology related to complex adaptive systems, when it comes to the consideration of systems composed of humans, their complexity increases significantly \citep{Karwowski2012}. This might be due to a multiplicity of factors related to human behavior and interactions among them, such as culture, attitudes, and cognitive abilities. Regarding preparedness with respect to hazardous events, the functioning of soft factors and their interrelations with human factors might be even more challenging to capture in models, as situations with less predictable and more complex patterns in individual behavior need to be covered \citep{Poletti2009,Funk2009,Reluga2010,DelValle2005}. 

The development of sophisticated models of adaptive human systems which capture the dynamics related to pandemics -- with a particular focus on the soft-facts related to resilience -- appears to be highly relevant. Such models might prov to be particularly useful when used to evaluate means to guide societies toward self-organization in order to increase their resilience.

\section{Final remarks}

This paper highlights some issues which emerge from pandemics and accompanying infodemics: Delayed effects of pandemics and factors influencing these effects, and the preparedness of adaptive societies. Understanding the dynamics related to pandemics and infodemics is an indisputable precondition for efficient and effective  decision-making. The discussion provided in this paper hopefully gives an impulse to action towards research related to hazardous events that puts more emphasis on the emerging behavioral, social, and economic dynamics.

%
\section*{Conflict of interest}
The author declares that he has no conflict of interest.

\bibliographystyle{spbasic}      
\bibliography{bibliography}   

\end{document}